\documentclass[conference]{IEEEtran}
\usepackage{graphicx}
\usepackage{cite}
\usepackage{amsmath,amssymb,amsfonts}
\usepackage{algorithmic}
\usepackage{textcomp}
\usepackage{xcolor}
\usepackage{booktabs}
\usepackage{svg}
\usepackage{xspace}
\usepackage[acronym]{glossaries}
\makeglossaries

\usepackage{booktabs} 
\usepackage{url}
\usepackage{color}
\usepackage{multirow}
\usepackage{enumitem}
\usepackage{array}
\usepackage[nolist,nohyperlinks]{acronym}
\usepackage{footnote}
\makesavenoteenv{tabular}

\usepackage{fancyhdr}
\def\BibTeX{{\rm B\kern-.05em{\sc i\kern-.025em b}\kern-.08em
    T\kern-.1667em\lower.7ex\hbox{E}\kern-.125emX}}

\acrodef{mIoU}{mean Intersection over Union}
\acrodef{DL}{deep learning}
\acrodef{IOU}[IoU]{Intersection over Union}
\acrodef{ROI}{region of interest}
\acrodef{ML}{machine learning}
\acrodef{MICCAI}{Medical Image Computing and Computer Assisted Intervention Society}
\acrodef{MIS}{Minimally Invasive Surgery}
\acrodef{CNN}{convolutional neural network}
\acrodef{DSC}{dice coefficient}

\begin{document}
\title{Exploring Deep Learning Methods for Real-Time Surgical Instrument Segmentation in Laparoscopy}

\author{\IEEEauthorblockN{Debesh Jha\IEEEauthorrefmark{1}\IEEEauthorrefmark{2}, 
Sharib Ali\IEEEauthorrefmark{4}\IEEEauthorrefmark{9},
Nikhil Kumar Tomar\IEEEauthorrefmark{1},
Michael A. Riegler\IEEEauthorrefmark{1}\IEEEauthorrefmark{2},\\
Dag Johansen\IEEEauthorrefmark{2},
H{\aa}vard D. Johansen\IEEEauthorrefmark{2},
P{\aa}l Halvorsen\IEEEauthorrefmark{1}\IEEEauthorrefmark{4}}

\IEEEauthorblockA{\IEEEauthorrefmark{1}SimulaMet, Norway \ \ \ \ \
\IEEEauthorrefmark{2}UiT The Arctic University of Norway, Norway \ \ \ \ \
\IEEEauthorrefmark{3}Oslo Metropolitan University, Norway \\ 
\IEEEauthorrefmark{4}Institute of Biomedical Engineering, University of Oxford, UK\ \ \ \
\IEEEauthorrefmark{9}NIHR Oxford Biomedical Research Centre, UK\\
}
}

\maketitle
\thispagestyle{fancy}
\begin{abstract}
\ac{MIS} is a surgical intervention used to examine the organs inside the abdomen and has been widely used due to its effectiveness over open surgery. Due to the hardware improvements such as high definition cameras, this procedure has significantly improved and new software methods have demonstrated potential for computer-assisted procedures. However, there exists challenges and requirement to improve detection and tracking of the position of the instruments during these surgical procedures. To this end, we evaluate and compare some popular deep learning methods that can be explored for the automated segmentation of surgical instruments in laparoscopy, an important step towards tool tracking. Our experimental results exhibit that the Dual decoder attention network (DDANet) produces a superior result compared to other recent deep learning methods. DDANet yields a Dice coefficient of 0.8739 and mean intersection-over-union of 0.8183 for the Robust Medical Instrument Segmentation (ROBUST-MIS) Challenge 2019 dataset, at a real-time speed of 101.36 frames-per-second that is critical for such procedures. 

\end{abstract}

\begin{IEEEkeywords}
Deep learning, segmentation, minimally invasive surgery, surgical instruments,  laparoscopy, real-time 
\end{IEEEkeywords}

\section{Introduction}
\label{sec:introduction}

\acf{MIS} is one of the most successful surgical intervention methods performed by inserting the surgical instruments inside the body of the patient through one or more small incisions~\cite{bartoli2012computer}. Laparoscopy is one of the types of \ac{MIS}. It is a procedure with less pain, minor cuts, reduced blood loss, shorter hospital stay, and fewer complications than traditional open surgery, i.e., overcoming limitations of the traditional laparoscopic surgery~\cite{azqueta2020segmentation}. 
However, moving of the surgical instrument is limited due to restricted operating space, which causes problems such as limited field of view of the endoscope, reduction in surgeon's dexterity, and lacks perception of force feedback~\cite{qiu2019real}. One of the solutions to overcome these difficulties could be specialized surgical training of the surgeon for better visualization~\cite{zahiri2017application}. Another solution could be the initiative to build computer vision based applications to assist surgeons. A computer and robotic-assisted surgical system can enhance the capability of the surgeons~\cite{cleary2010image} and be helpful for decision making during the surgery~\cite{bodenstedt2018comparative}. However, to make this possible, the challenge of understanding the spatial relationships between surgical instruments, cameras, and anatomy for the patient needs to be solved~\cite{pakhomov2019deep}.  

Segmentation of surgical instruments is a difficult task because it has challenging conditions such as blood, smoke, reflection, and motion artifacts. During surgery, a strong lightning condition is required for proper visualization. The intense lights can lead to non-negligible specularity on the tissue and the surgical instrument. Due to such specular reflections, the surface color can be confused with other similar colored surfaces having the same hue irrespective of having lower saturation~\cite{liu2015saturation}. In a surgical procedure, specular reflections cause the surgical instrument to appear white. Additionally, shadows can occur in the video frames due to the change in the illumination angle and movement of the surgical tool during surgery~\cite{ni2020attention}.  

Previous work has targeted instrument segmentation, detection, and tracking in endoscopic video images. However, they fail on challenging images such as images with the presence of specularity, blood, smoke, and motion artifacts~\cite{ross2021comparative}. This study aims to find efficient computer vision methods for surgical instrument segmentation from laparoscopic images. We evaluate several recent state-of-the-art deep learning approaches with different backbones using the ROBUST-MIS challenge dataset acquired from real world clinical surgical procedures. 


    
    
    

\section{Related Work}
\label{sec:related work}

Surgical computer vision is evolving as promising techniques to segment and track the instruments using endoscopic images are emerging~\cite{bodenstedt2018comparative,ross2021comparative}. Bodenstedt et al.~\cite{bodenstedt2018comparative} organized ``EndoVis 2015 Instrument sub-challenge\footnote{\url{https://endovissub-instrument.grand-challenge.org/EndoVisSub-Instrument/}}", a part of Endoscopic vision (EndoVis) challenge\footnote{\url{https://endovis.grand-challenge.org/}}.
In 2017, a similar challenge, ``Robotic Instrument Segmentation Sub-Challenge\footnote{\url{https://endovissub2017-roboticinstrumentsegmentation.grand-challenge.org/}}" was organized at the same platform. At this time, the organizers increased the number of tasks to three. The participants were asked to participate in a binary segmentation task, a parts based segmentation task, and an instrument type segmentation task. Both of these challenges provided sufficient insights on the instrument segmentation. However, the insights gained on the robustness and generalization capability of the deep learning methods were limited~\cite{bodenstedt2018comparative,ross2021comparative}.

A similar Challenge ``Robust Medical Instrument Segmentation Challenge 2019\footnote{\url{https://robustmis2019.grand-challenge.org/}}" was organized by Ro{\ss} et al.~\cite{ross2021comparative}. The challenge aimed to develop a potential method for minimally invasive surgery to track the surgical instruments in the abdomen. Most of these methods included Mask-RCNN~\cite{he2017mask} based instance segmentation for multi-class segmentation and for binary segmentation. Participants also explored methods such as OR-UNet~\cite{isensee2020or},  DeepLabV3+\cite{chen2018encoder}, U-Net~\cite{ronneberger2015u} and RASNet~\cite{ni2019rasnet}. The best performing methods for the binary instrument segmentation were OR-UNet and the DeepLabv3+ with pre-trained ImageNet encoders. Ceron et al.~\cite{arXiv:Juan2021} recently proposed an attention-based segmentation for MIS instruments on the same datasets that achieved state-of-the-art performance at nearly 45 frames-per-second. In addition to these challenges, there are works on segmentation~\cite{lee2019segmentation} and identification of surgical instruments in laparoscopy~\cite{kletz2019identifying}. We aim to segment surgical instruments from laparoscopy in real-time with well-established deep learning methods in this work.

\section{Material}
\label{sec:Material}
\subsection{Dataset}
ROBUST-MIS is part of the Heidelberg Colorectal (HeiCo) dataset~\cite{maier2021heidelberg,ross2021comparative}, which includes a total of $5983$ images with their respective manually annotated segmentation masks. These training images are obtained from $16$ different surgeries with two different types of procedures (prokto and rectum, $8$ surgeries each). The prokto sub-folder consists of $2943$ images, and the rectum sub-folder contains $3040$ images. Our experiments were performed on the training dataset provided by the ROBUST-MIS Challenge organizers.




\subsection{Data Pre-processing}
We have only considered the image dataset for training to reduce the computational cost. The original size of the images is 960$\times$540. We split the dataset into (80-10-10) for training, validation, and testing, respectively. The training dataset was augmented and resized. The total number of images of the training dataset is 4787, and the validation and test dataset contains 598 samples each.  


\subsection{Experimental setup and configuration}
All of the experiments were implemented using the PyTorch framework except ResUNet~\cite{zhang2018road} and ResUNet++~\cite{jha2019resunet++} that were implemented with TensorFlow~\cite{abadi2016tensorflow}. The experiments such as ResUNet\cite{zhang2018road}, ResUNet++~\cite{jha2019resunet++}, ColonSegNet~\cite{jha2021real}, DDANet~\cite{tomar2020ddanet} were run on the Experimental Infrastructure for Exploration of Exascale Computing (eX3), NVIDIA DGX-2 machine. These models were trained using offline augmentation. The models such as UNet~\cite{ronneberger2015u}, FCN8~\cite{long2015fully}, PSPNet~\cite{zhao2017pyramid}, DeepLabv3+~\cite{chen2018encoder} with MobileNetv2~\cite{howard2017mobilenets} and ResNet50~\cite{he2016deep} backbones were trained on on NVIDIA  Quadro  RTX  6000 and an online augmentation strategy was adopted.  This is because the experiments were run on two different GPUs. Adam optimizer was adopted for all experiments. The other hyperparameters were tuned based on the empirical evaluation. 

\begin{table*}[!t]
\centering
 \caption{Evaluation of surgical instruments segmentation task on the  ROBUST-MIS Challenge 2019 dataset}
    \label{table:quantitativeresults}
     \setlength\tabcolsep{6px}
\begin{tabular}{l|l|c|c|c|c|c|c|c}
\toprule
\textbf{Method}& \textbf{Backbone}& \textbf{DSC} & \textbf{mIoU} & \textbf{Recall} & \textbf{Precision}& \textbf{F2} & \textbf{Accuracy} & \textbf{FPS}\\
\midrule

UNet (MICCAI 2015)~\cite{ronneberger2015u} &ResNet34~\cite{he2016deep} &0.4214 &0.3318 &0.5071 &0.4224 & 0.4502 & 0.9294 &35.00\\

FCN8 (CVPR 2015)~\cite{long2015fully}  & - &0.7774 &0.6825 &0.7737 &0.8481 &0.7600 &0.9817 & 24.91\\

ResUNet (GRSL 2017)~\cite{zhang2018road} &- & 0.7097 & 0.6193 & 0.8056 & 0.7627 & 0.7111 & 0.9755 & 10.40\\ 

PSPNet (CVPR 2017)~\cite{zhao2017pyramid} &- & 0.8110 & 0.7235 & 0.7677 & 0.9278 & 0.7715 & 0.9855 & 16.80\\

Deeplabv3+ (ECCV 2018)~\cite{chen2018encoder} & MobileNetv2~\cite{howard2017mobilenets} & 0.8497 & 0.7780 & 0.8338 & 0.9246 & 0.8281 & 0.9882 & 32.50\\ 

Deeplabv3+  (ECCV  2018)~\cite{chen2018encoder} & ResNet50~\cite{he2016deep} &0.8582 & 0.7895 & 0.8397 & \textbf{0.9301} & 0.8362 & 0.9889 & 27.90\\

ResUNet++ (ISM 2019)~\cite{jha2019resunet++} &- & 0.8252 & 0.7576 & 0.8460 & 0.8903 & 0.8319 & 0.9811 & 8.13\\ 

ColonSegNet (IEEE Access 2021)~\cite{jha2021real} &- & 0.8495 & 0.7943 & \textbf{0.8899} & 0.8871 & 0.8443 & \textbf{0.9898} & \textbf{185.54}\\ 

DDANet (ICPRW 2020)~\cite{tomar2020ddanet} &- & \textbf{0.8739} & \textbf{0.8183} & 0.8703 & 0.9348 & \textbf{0.8613} & \textbf{0.9897} & 101.36\\

\bottomrule
\end{tabular}
\end{table*} 

\begin{figure*} [!t]
    \centering
    \includegraphics[width = \linewidth]{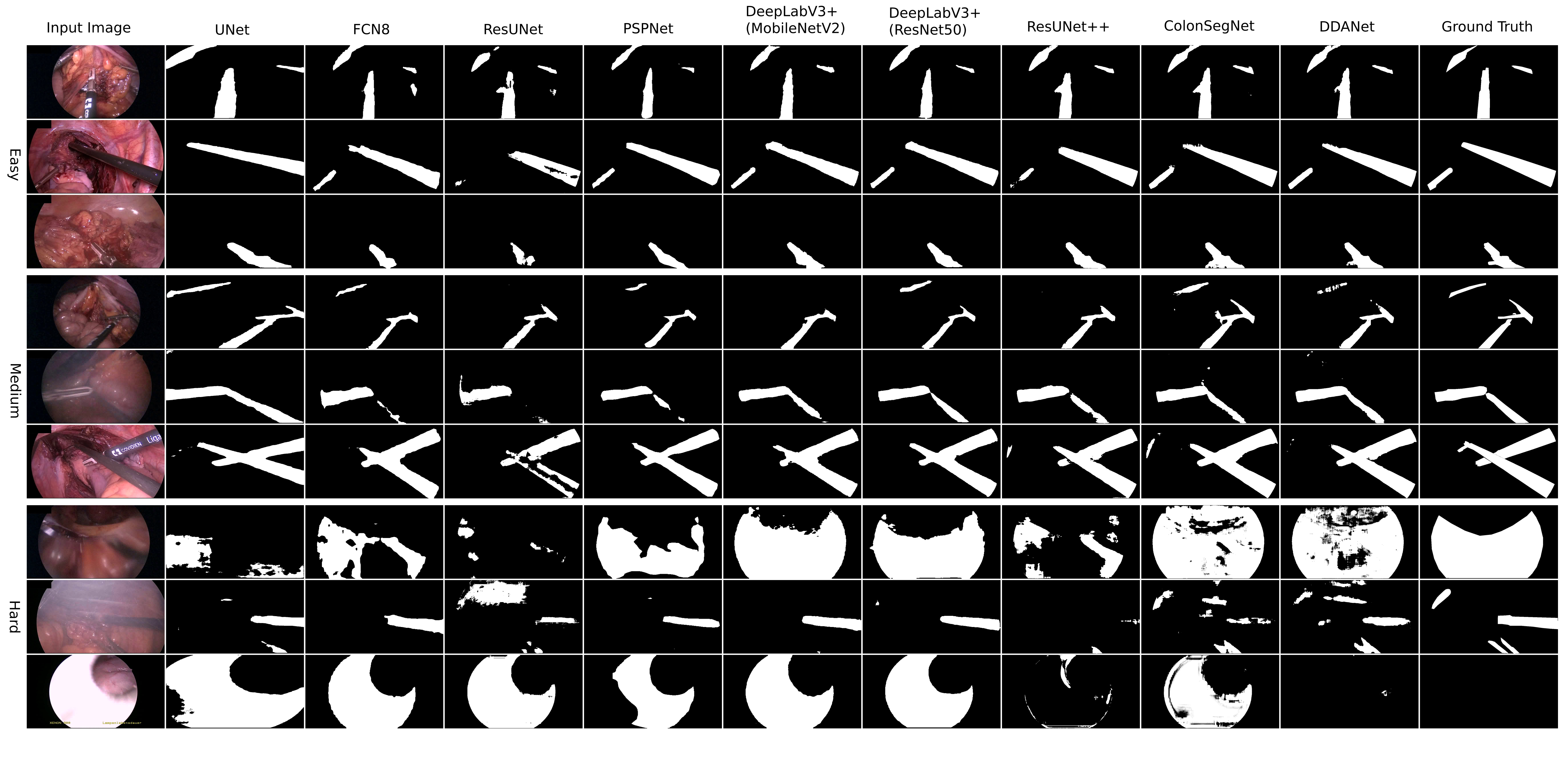}
    \caption{Qualitative results comparison of nine deep learning methods on easy cases, medium cases, and hard cases}
    \label{fig:differentcases}
\end{figure*}

\section{Methodology}
\label{sec:methods}

The goal of a segmentation method is to label each pixel of the image into either a surgical instrument or background, i.e., the output of the model is a pixel-by-pixel array that determines the class of each pixel. We have compared nine different state-of-the-art recent deep learning methods such as UNet~\cite{ronneberger2015u}, FCN8~\cite{long2015fully}, ResUNet~\cite{zhang2018road}, ResUNet++~\cite{jha2019resunet++}, DDANet~\cite{tomar2020ddanet}, ColonSegNet~\cite{jha2021real}, PSPNet~\cite{zhao2017pyramid}, and DeepLabv3+~\cite{chen2018encoder} with ResNet50~\cite{he2016deep} and MobileNetv2~\cite{howard2017mobilenets} backbones. The main motivation behind choosing these networks is that they have shown state-of-the-art results for various medical image segmentation tasks. A detailed explanation about the overall architectures, idea behind using each component in the architecture, advantage of the blocks used, and performance on different imaging modalities can be found in their respective papers. 

\section{Results}
\label{sec:results}

In this section, we present the experimental results. We report the result in Table~\ref{table:quantitativeresults} using standard metrics used for computer vision tasks. From Table~\ref{table:quantitativeresults}, we observe that DDANet produces the highest \ac{DSC} of 0.8739, \ac{mIoU} of 0.8183, recall of 0.8703, precision of 0.9348, F2 of 0.8613, accuracy of 0.9897 and real-time speed of 101.36. The other most competitive method was ColonsegNet, as it has the competitive DSC, mIoU, precision, and F2. ColonsegNet also has the highest recall of  0.8899 and highest FPS of 185.54. The precision for DeepLabv3+ with ResNet50 backbone is the best. From the quantitative result analysis, we can see that there is a significant improvement in deep learning methods as we compare methods from 2015 to 2021. The current methods are efficient in terms of both accuracy and speed.  

To show the effectiveness of the different models, we have shown the qualitative analysis results of nine different deep learning methods on easy, medium, and hard images as shown in Figure~\ref{fig:differentcases}. From the qualitative analysis, we can see that all the models perform satisfactorily on the easy cases, although it has one instrument or multiple instruments (Figure~\ref{fig:differentcases}). For the medium cases, the deep learning methods such as Deeplabv3, ColonsegNet, and DDANet are performing satisfactorily. However, for ColonsegNet and DDANet, we can also observe over-segmentation. For the hard cases, Deeplabv3 with both the backbones shows better segmentation results for some scenarios, whereas ColonSegNet and DDANet show over-segmentation. For some challenging scenarios, DDANet shows promising results (please refer to the last image of Figure~\ref{fig:differentcases}). Overall, the preferred choice is DDANet, although it shows over-segmentation for some scenarios. The strength is also because of its capability to segment surgical instruments in real-time with a high FPS that was not achieved by other methods such as DeepLabv3+ with different backbones.  

\section{Discussion}
\label{sec:discussion}
MIS is a progressive field where computer-assisted techniques are being iterated and developed for clinical use. While, segmentation of surgical instruments can play a vital role in tracking and a step forward towards automated systems, the real-time requirement hinders the clinical applicability of many recent deep learning methods. We have presented different recent segmentation methods and identified efficient architectures for the pixel-wise segmentation of surgical instruments on the laparoscopy dataset. As observed from Table~\ref{table:quantitativeresults}, DDANet architecture provides the highest metric performances over other methods and has real-time capability (101.36 FPS). As evident from Figure~\ref{fig:differentcases}, DDANet is a robust encoder-decoder network that can capture sharper surgical instrument boundaries by continuously retrieving the spatial information. The encoder branch in the DDANet helps boost performance by generating the attention map (spatial attention), which is used in the segmentation branch to improve the feature representation. Since the ground truth is not available for the test set, we can not directly compare the results with the challenge participants. However, our results on the validation set show promise and reliance of the DDANet. Furthermore, we also explored the ability of models on a low-end NVIDIA 1060 Ti GPU (6 GB). For this, we achieved an FPS of 120.05 on ColonSegNet and 95.05 on DDANet which suggests the effectiveness of these models. 


We explicitly choose to use a pre-trained encoder with the DeepLabv3+ because we hypothesize that the ImageNet pre-trained network will improve performance. Although it provided a promising \ac{DSC} of 0.8582, the methods such as ColonSegNet and DDANet showed competitive or better results without learning from any pre-trained encoders. The main reason for higher FPS for these two models is the lightweight network architecture that making the model efficient for real-time segmentation facilitating clinical deployment. The detailed analysis of the qualitative results (Figure~\ref{fig:differentcases}) show that the best methods are robust to samples with specularity, blood, and smoke, although it shows some areas with over segmentation. This is also evident from the dice score of the best performing method. However, the challenge still exists with the images having moving instruments and the over/under segmentation problem with the hard cases (see bottom rows, Figure~\ref{fig:differentcases}) that need to be addressed in further study. 


%
%
\section{Conclusion}
\label{sec:conclusion}
This paper explores several deep learning methods with different backbones for the segmentation of surgical instruments in laparoscopy. A comparison of different state-of-the-art methods shows that DDANet is the most efficient method providing a strong baseline on the ROBUST-MIS challenge 2019 dataset. DDANet provides the most accurate segmentation of the surgical instruments with a real-time speed of 101.36 FPS. The provided comparison shows the strength of simple and small networks such as ColonSegNet and DDANet to achieve real-time performance with competitive scores compared to that of complex and hybrid networks. In the domains where real-time inference is vital, exploring such networks are important to accelerate both research and clinical applicability of models. The achieved results also highlight the potential for the extension of the work to achieve higher accuracy by incorporating attention mechanisms or use of skip-connections during the feature flow but maintaining the speed at the required real-time performances, e.g., 60 FPS for most devices. We plan to investigate multi-instance segmentation and a composite instrument tracking technique for future work.

\section*{Acknowledgement}
D. Jha is funded by  Research Council of Norway project number 263248 (Privaton). The computations in this paper were performed on equipment provided by the Experimental Infrastructure for Exploration of Exascale Computing (eX3), which is financially supported by the Research Council of Norway under contract 270053. 

\bibliographystyle{IEEEtran}
\bibliography{references} 

\begin{thebibliography}{10}
\providecommand{\url}[1]{#1}
\csname url@samestyle\endcsname
\providecommand{\newblock}{\relax}
\providecommand{\bibinfo}[2]{#2}
\providecommand{\BIBentrySTDinterwordspacing}{\spaceskip=0pt\relax}
\providecommand{\BIBentryALTinterwordstretchfactor}{4}
\providecommand{\BIBentryALTinterwordspacing}{\spaceskip=\fontdimen2\font plus
\BIBentryALTinterwordstretchfactor\fontdimen3\font minus
  \fontdimen4\font\relax}
\providecommand{\BIBforeignlanguage}[2]{{%
\expandafter\ifx\csname l@#1\endcsname\relax
\typeout{** WARNING: IEEEtran.bst: No hyphenation pattern has been}%
\typeout{** loaded for the language `#1'. Using the pattern for}%
\typeout{** the default language instead.}%
\else
\language=\csname l@#1\endcsname
\fi
#2}}
\providecommand{\BIBdecl}{\relax}
\BIBdecl

\bibitem{bartoli2012computer}
A.~Bartoli, T.~Collins, N.~Bourdel, and M.~Canis, ``Computer assisted minimally
  invasive surgery: is medical computer vision the answer to improving
  laparosurgery?'' \emph{Medical Hypotheses}, vol.~79, no.~6, pp. 858--863,
  2012.

\bibitem{azqueta2020segmentation}
I.~Azqueta-Gavaldon, F.~Fr{\"o}hlich, K.~Strobl, and R.~Triebel, ``Segmentation
  of surgical instruments for minimally-invasive robot-assisted procedures
  using generative deep neural networks,'' \emph{arXiv preprint
  arXiv:2006.03486}, 2020.

\bibitem{qiu2019real}
L.~Qiu, C.~Li, and H.~Ren, ``Real-time surgical instrument tracking in
  robot-assisted surgery using multi-domain convolutional neural network,''
  \emph{Healthcare Technology Letters}, vol.~6, no.~6, pp. 159--164, 2019.

\bibitem{zahiri2017application}
M.~Zahiri, \emph{Application of computer vision in surgical training and
  surgical robotics}.\hskip 1em plus 0.5em minus 0.4em\relax The University of
  Nebraska-Lincoln, 2017.

\bibitem{cleary2010image}
K.~Cleary and T.~M. Peters, ``Image-guided interventions: technology review and
  clinical applications,'' \emph{Annual review of biomedical engineering},
  vol.~12, pp. 119--142, 2010.

\bibitem{bodenstedt2018comparative}
S.~Bodenstedt \emph{et~al.}, ``Comparative evaluation of instrument
  segmentation and tracking methods in minimally invasive surgery,''
  \emph{arXiv preprint arXiv:1805.02475}, 2018.

\bibitem{pakhomov2019deep}
D.~Pakhomov, V.~Premachandran, M.~Allan, M.~Azizian, and N.~Navab, ``Deep
  residual learning for instrument segmentation in robotic surgery,'' in
  \emph{Proc. of International Workshop on Machine Learning in Medical
  Imaging}, 2019, pp. 566--573.

\bibitem{liu2015saturation}
Y.~Liu, Z.~Yuan, N.~Zheng, and Y.~Wu, ``Saturation-preserving specular
  reflection separation,'' in \emph{Proc. of the IEEE Conference on Computer
  Vision and Pattern Recognition (CVPR)}, 2015, pp. 3725--3733.

\bibitem{ni2020attention}
Z.-L. Ni, G.-B. Bian, Z.-G. Hou, X.-H. Zhou, X.-L. Xie, and Z.~Li,
  ``Attention-guided lightweight network for real-time segmentation of robotic
  surgical instruments,'' in \emph{Proc. of International Conference on
  Robotics and Automation (ICRA)}, 2020, pp. 9939--9945.

\bibitem{ross2021comparative}
T.~Ro{\ss} \emph{et~al.}, ``Comparative validation of multi-instance instrument
  segmentation in endoscopy: Results of the robust-mis 2019 challenge,''
  \emph{Medical image analysis}, vol.~70, p. 101920, 2021.

\bibitem{he2017mask}
K.~He, G.~Gkioxari, P.~Doll{\'a}r, and R.~Girshick, ``Mask r-cnn,'' in
  \emph{Proc. of the IEEE international conference on computer vision (ICCV)},
  2017, pp. 2961--2969.

\bibitem{isensee2020or}
F.~Isensee and K.~H. Maier-Hein, ``Or-unet: an optimized robust residual u-net
  for instrument segmentation in endoscopic images,'' \emph{arXiv preprint
  arXiv:2004.12668}, 2020.

\bibitem{chen2018encoder}
L.-C. Chen, Y.~Zhu, G.~Papandreou, F.~Schroff, and H.~Adam, ``Encoder-decoder
  with atrous separable convolution for semantic image segmentation,'' in
  \emph{Proc. of the European conference on computer vision (ECCV)}, 2018, pp.
  801--818.

\bibitem{ronneberger2015u}
O.~Ronneberger, P.~Fischer, and T.~Brox, ``U-net: Convolutional networks for
  biomedical image segmentation,'' in \emph{Proc. of Medical Image Computing
  and Computer Assisted Intervention (MICCAI)}, 2015, pp. 234--241.

\bibitem{ni2019rasnet}
Z.-L. Ni, G.-B. Bian, X.-L. Xie, Z.-G. Hou, X.-H. Zhou, and Y.-J. Zhou,
  ``Rasnet: segmentation for tracking surgical instruments in surgical videos
  using refined attention segmentation network,'' in \emph{2019 41st Annual
  International Conference of the IEEE Engineering in Medicine and Biology
  Society (EMBC)}, 2019, pp. 5735--5738.

\bibitem{arXiv:Juan2021}
J.~C.~A. C\'{e}ron, L.~Chang, G.~Ochoa{-}Ruiz, and S.~Ali, ``Assessing
  {YOLACT++} for real time and robust instance segmentation of medical
  instruments in endoscopic procedures,'' \emph{arXiv preprint
  arXiv:2103.15997}, 2021.

\bibitem{lee2019segmentation}
E.-J. Lee, W.~Plishker, X.~Liu, T.~Kane, S.~S. Bhattacharyya, and R.~Shekhar,
  ``Segmentation of surgical instruments in laparoscopic videos: training
  dataset generation and deep-learning-based framework,'' in \emph{Proc. of
  Medical Imaging 2019: Image-Guided Procedures, Robotic Interventions, and
  Modeling}, vol. 10951, 2019, p. 109511T.

\bibitem{kletz2019identifying}
S.~Kletz, K.~Schoeffmann, J.~Benois-Pineau, and H.~Husslein, ``Identifying
  surgical instruments in laparoscopy using deep learning instance
  segmentation,'' in \emph{Proc. of International Conference on Content-Based
  Multimedia Indexing (CBMI)}, 2019, pp. 1--6.

\bibitem{maier2021heidelberg}
L.~Maier-Hein \emph{et~al.}, ``Heidelberg colorectal data set for surgical data
  science in the sensor operating room,'' \emph{Scientific Data}, vol.~8,
  no.~1, pp. 1--11, 2021.

\bibitem{zhang2018road}
Z.~Zhang, Q.~Liu, and Y.~Wang, ``Road extraction by deep residual u-net,''
  \emph{IEEE Geoscience and Remote Sensing Letters}, vol.~15, no.~5, pp.
  749--753, 2018.

\bibitem{jha2019resunet++}
D.~Jha \emph{et~al.}, ``Resunet++: An advanced architecture for medical image
  segmentation,'' in \emph{Proc. of International Symposium on Multimedia
  (ISM)}, 2019, pp. 225--2255.

\bibitem{abadi2016tensorflow}
M.~Abadi \emph{et~al.}, ``Tensorflow: A system for large-scale machine
  learning,'' in \emph{Proc. of Proceeding of USENIX Symposium on Operating
  Systems Design and Implementation (OSDI)}, 2016, pp. 265--283.

\bibitem{jha2021real}
D.~Jha \emph{et~al.}, ``Real-time polyp detection, localization and
  segmentation in colonoscopy using deep learning,'' \emph{Ieee Access},
  vol.~9, pp. 40\,496--40\,510, 2021.

\bibitem{tomar2020ddanet}
N.~K. Tomar \emph{et~al.}, ``{DDANet}: Dual decoder attention network for
  automatic polyp segmentation,'' in \emph{Proc. of ICPRW}, 2020.

\bibitem{long2015fully}
J.~Long, E.~Shelhamer, and T.~Darrell, ``Fully convolutional networks for
  semantic segmentation,'' in \emph{Proc. of the IEEE conference on computer
  vision and pattern recognition (CVPR)}, 2015, pp. 3431--3440.

\bibitem{zhao2017pyramid}
H.~Zhao, J.~Shi, X.~Qi, X.~Wang, and J.~Jia, ``Pyramid scene parsing network,''
  in \emph{Proc. of the IEEE conference on computer vision and pattern
  recognition (CVPR)}, 2017, pp. 2881--2890.

\bibitem{howard2017mobilenets}
A.~G. Howard \emph{et~al.}, ``Mobilenets: Efficient convolutional neural
  networks for mobile vision applications,'' \emph{arXiv preprint
  arXiv:1704.04861}, 2017.

\bibitem{he2016deep}
K.~He, X.~Zhang, S.~Ren, and J.~Sun, ``Deep residual learning for image
  recognition,'' in \emph{Proc. of the IEEE conference on computer vision and
  pattern recognition (CVPR)}, 2016, pp. 770--778.

\end{thebibliography}
\end{document}